\documentclass[%
 preprint,
 amsmath,amssymb,
 aps,
]{revtex4-2}

\usepackage{graphicx}
\usepackage{dcolumn}
\usepackage{bm}
\usepackage{nicefrac}
\usepackage{xcolor}

\begin{document}


\title{
Charge Transport and Defects in Sulfur-Deficient Chalcogenide Perovskite BaZrS$_3$
}
\author{Garima Aggarwal$^{a\dag}$}
\email{garima.agrawalsk@gmail.com\\ $\dag$Now at ReNew Private Limited, Gurgaon - India\\$^{\ddag}$These authors contributed equally}

\author{Adeem Saeed Mirza$^{b\ddag}$, Stefania Riva$^{c\ddag}$}
\author{Corrado Comparotto$^a$, Robert J. W. Frost$^d$, Soham Mukherjee$^c$, Monica Morales-Masis$^b$, H$\mathring{a}$kan Rensmo$^c$, and Jonathan Staaf Scragg$^a$}

\affiliation{$^a$Division of Solar Cell Technology, Department of Materials Science and Engineering, Uppsala University, Uppsala 75237, Sweden\\
$^b$ MESA+ Institute for Nanotechnology, University of Twente, Enschede 7500 AE, The Netherlands\\
$^c$ Molecular and Condensed Matter Physics, Department of Physics and Astronomy, Uppsala University Box 516, SE-75120 Uppsala, Sweden\\
$^d$ Division of Applied Nuclear Physics, Uppsala University, Uppsala 75237, Sweden}

\begin{abstract}
Exploring the conduction mechanism in the chalcogenide perovskite BaZrS$_3$ is of significant interest due to its potential suitability as a top absorber layer in silicon-based tandem solar cells and other optoelectronic applications. Theoretical and experimental studies anticipate native ambipolar doping in BaZrS$_3$, although experimental validation remains limited. This study reveals a transition from highly insulating behavior to n-type conductivity in BaZrS$_3$ through annealing in an S-poor environment. BaZrS$_3$ thin films are synthesized $\emph{via}$ a two step process: co-sputtering of Ba-Zr followed by sulfurization at 600 $^{\circ}$C, and subsequent annealing in high vacuum. UV-Vis measurement reveal a red-shift in the absorption edge concurrent with sample color darkening after annealing. The increase in defect density with vacuum annealing, coupled with low activation energy and n-type character of defects, strongly suggests that sulfur vacancies (V$_{\mathrm{S}}$) are responsible, in agreement with theoretical predictions. The shift of the Fermi level towards conduction band minimum, quantified by Hard X-ray Photoelectron Spectroscopy (Ga K$\alpha$, 9.25 keV), further corroborates the induced n-type of conductivity in annealed samples. Our findings indicate that vacuum annealing induces V$_{\mathrm{S}}$ defects that dominate the charge transport, thereby making BaZrS$_3$ an n-type semiconductor under S-poor conditions. This study offers crucial insights into understanding the defect properties of BaZrS$_3$, facilitating further improvements for its use in solar cell applications.

\end{abstract}

\maketitle


\section{Introduction}

Perovskite chalcogenides are emerging as highly interesting optoelectronic materials \cite{sun2015chalcogenide, sopiha2022chalcogenide, raj2023chalcogenide, zhang2023prediction}. These materials exhibit a bandgap ranging from 1.1 to 2.0 eV with an absorption coefficient greater than 10$^5$ cm$^{-1}$ in the visible range of the solar spectrum. Perovskite chalcogenides, as the name implies, crystallize in the perovskite structure characterized by the chemical formula ABX$_3$, where chalcogens (often S or Se) occupy the anion 'X' site. Site `A' typically contains +2 charged cations (Ba, Sr, Ca), while site `B' contains +4 charged cations (Zr, Hf), compensating the -2 charged chalcogen anions. Several perovskite chalcogenide materials are being explored theoretically as well as experimentally depending on the elemental composition. Among these, BaZrS$_3$ has gained significant attention due to its high stability, the abundance of its constituent elements, and its near-optimal bandgap for a top cell active layer material in silicon-based tandem solar cells \cite{ye2022time, pal2024numerical, vincent2024emerging, mukherjee2023interplay, agarwal2023moderate, wei2020realization, comparotto2022synthesis, ramanandan2023understanding, zilevu2022solution, sadeghi2021making, filippone2020discovery}.

\begin{figure}
\centering
\includegraphics[width=0.35\textwidth]{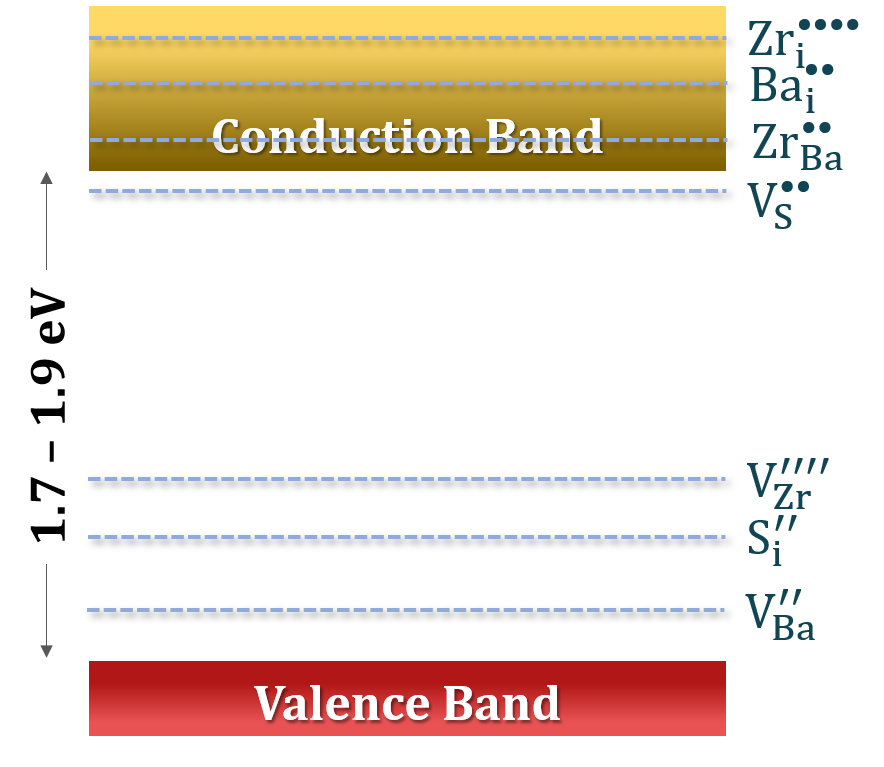}
\caption{A summarized band diagram schematic of BaZrS$_3$ from theoretical studies, illustrating point defect levels: acceptor-type defects with a 0/- ionization level close to the VBM and donor-type defects with a 0/+ ionization level close to the CBM (all defects are represented in Kr$\mathrm{\ddot{o}}$ger--Vink notation) \cite{wu2021defect, thakur2023recent, wu2021defect, osei2021examining}.}
\label{defects}
\end{figure}

In the literature, most studies focus on process development and optical properties, with limited attention paid to charge transport \cite{turnley2022solution, yang2022low, comparotto2022synthesis, ramanandan2023understanding, agarwal2023moderate, yang2023low, comparotto2020chalcogenide, pradhan2023synthesis}. However, understanding defect characteristics is key for establishing good performance and controllability in device applications, as well as for avoiding unwanted recombination processes. A number of theoretical studies report BaZrS$_3$ to exhibit intrinsic p-type or n-type behavior, depending on the thermodynamics of growth conditions \cite{wu2021defect, thakur2023recent, wu2021defect, osei2021examining}. However, to the best of our knowledge, there is limited experimental evidence available to support this claim. Theoretical studies propose several point defects to be present in BaZrS$_3$: including both acceptor as well as donor type of defects. A summary of these defect levels is represented in Fig.~\ref{defects}, based on the reported literature. 
The defects are represented in Kr$\ddot{o}$ger Vink notation. We have selectively presented defects with ionization energy of a donor (acceptor) type defect being close to the conduction band minimum (valence band maximum) as they are either resonant defects (within CB or VB) or require minimum energy for ionization at room temperature. Here, V$_{\mathrm{S}}$, Zr$_{\mathrm{i}}$, Ba$_{\mathrm{i}}$, Zr$_{\mathrm{Ba}}$ are donor type of defects, while S$_{\mathrm{i}}$, V$_{\mathrm{Zr}}$, and V$_{\mathrm{Ba}}$ are acceptor type of defects. The charge state transition level (CSTL) of the former (0/+) and the latter (0/-) that are close to CBM and VBM, respectively are presented with dashed lines in the figure. Other defects with high ionization energy (deep levels) are not included, as they will not contribute significantly to the conduction at room temperature. The type of conduction in BaZrS$_3$ will depend on which of the defect levels in Fig.~\ref{defects} dominate at room temperature. 

Typically, as-prepared samples of BaZrS$_3$ exhibit high resistivity, making it challenging to study the conduction mechanism. This behavior may result from poor film quality (crystallinity, morphology, phase purity, surface impurity) or/and the chemical composition being close to stoichiometry, i.e. low concentration of either intrinsic donor or acceptor type of defects. To achieve the desired conductivity type and carrier concentration, it is crucial to first improve the film quality, but potentially also modify the synthesis process, such as changing conductivity $\emph{via}$ either aliovalent doping on the cation/anion site or by varying the native cation/anion concentration \cite{han2023p, meng2016alloying}. 

In this article, we investigate charge transport in BaZrS$_3$ and its response to vacuum annealing. Initially, we characterize our as-prepared films using linear 4-probe resistivity measurements, revealing their highly insulating nature. The films are synthesized by the reaction of a metal precursor with a S-rich atmosphere, a process developed to achieve better crystallinity with minimal probability of impurity phase formation. Subsequently, these samples undergo a low-pressure heat treatment in the absence of S, inducing off-stoichiometry. Post-annealing, we observe an increase in the film conductivity, accompanied by a change in sample color from transparent orange to dark brown. UV-Vis spectroscopy reveal a red-shift in the bandgap after annealing. Energy-dispersive X-ray spectroscopy (EDS) analysis indicates that the surface of the samples become deficient in S after vacuum annealing, with the S atomic percentage (calculated as S/(Ba+Zr+S)) reducing by almost 2 \%. The samples exhibit n-type conductivity after annealing, with the electron concentration increasing with both annealing temperature and duration as observed in Hall measurement and Seebeck coefficient. This is further supported by depth-dependent X-ray Photoelectron Spectroscopy (XPS) using soft (Al K$\alpha$ -1.487 keV) and hard (Ga K$\alpha$ - 9.25 keV) X-rays. The results show a shift of the valence band maximum (VBM) towards higher binding energy relative to the Fermi level with an increase of annealing temperature and duration, with little to no interference of the VBM position from the film surface chemistry. Our findings highlight the possibility of controlling the nature of doping in BaZrS$_3$ thin films by optimizing processing conditions, paving way for its utilization in optoelectronic devices.

\subsection{Experimental Section}
The synthesis route involves sputtering of metal precursors; Ba and Zr with a Zr-interlayer (and protective layer SnS) followed by annealing in a S-rich atmosphere. This route is developed in our group to deposit phase pure BaZrS$_3$ films on variety of substrates [article under process]. A Kurt J Lesker system having six magnetron sources, each containing 3-inch targets, was used to deposit thin films. These were deposited at room temperature and a pressure of 5 mTorr in Ar atmosphere on a quartz substrate. The sulfurization of the sputtered films was performed in a home-built tubular furnace in Ar atmosphere. The sample and a sulfur source were placed separately in a graphite box and a molybdenum boat, respectively, inside a quartz tube. All films were sulfurized at around (7 $\pm$ 2) $\times$ 10$^{-1}$ mbar. The sample was maintained at 600 $^{\circ}$C and the S-source at 135 $^{\circ}$C. After thin film preparation, samples were diced into square sections (8 mm $\times$ 8 mm to 10 mm $\times$ 10 mm) such that they can be used in Van der Pauw configuration for Hall measurements. These diced samples were then annealed in a vacuum assisted tubular furnace to induce S-poor conditions. They are loaded into a ceramic boat inside a quartz tube. The temperature was varied from 300 to 600 $^{\circ}$ C, however no significant change in the film was observed below 500 $^{\circ}$C. We varied the duration from 0.7 to 2.1 h and observed a saturation in conductivity beyond 1.4 h (see SI). Based on this, we chose three sets of annealing conditions, which are used to name the samples using the notation T-t, where T is the temperature of annealing (in $^{\circ}$C) and t is the duration of the annealing (in h). We also compared the conductivity of samples with a varying Ba to Zr ratio; however, no significant difference was observed, and therefore it is not discussed here.

Transmittance and reflectance spectra were recorded in the wavelength range between 300 nm and 800 nm with a step resolution of 1.0 nm, employing a Lambda 950 double-beam UV/Vis/NIR spectrophotometer equipped with an integrating sphere and a Spectralon reflectance standard. Room temperature Hall measurements were carried out in Van der Pauw configuration and magnetic field of 1 T in an ezHEMS Hall setup by NanoMagnetics. Titanium-gold ohmic contacts were deposited on the corners of the BaZrS$_3$/quartz samples to facilitate electrical measurements. Seebeck measurements were performed in a home-built setup with samples placed on Peltier elements. The voltage difference caused by temperature differences was measured at the two ends of the sample. The slope of the temperature difference $\Delta$T versus the induced voltage V was calculated to obtain the Seebeck coefficient. Temperature-dependent conductivity measurements were performed in Van der Pauw configuration using a Quantum Design Physical Properties Measurement System (PPMS). A sweep of the magnetic field from +5 T to -5 T was done for resistivity measurements from 300 K to 10 K. The phase purity and crystallinity of the films were characterized $\emph{via}$ X-ray diffraction (Siemens D5000 and Empyrean) and no phase change was observed in films after annealing. Energy Dispersive Spectroscopy (EDS) (Zeiss LEO 1550 SEM with an EDAX EDS) was used to analyze the chemical composition of the samples, while Rutherford backscattering spectrometry (RBS) and Time-of-Flight Elastic recoil Detection Analysis were to provide depth-resolved compositional information. Both RBS and ToF-ERDA measurements were performed using the 5-MV NEC-5SDH-2 tandem accelerator at Uppsala University. RBS measurements were performed using primary beam of  2\,MeV \textsuperscript{4}He\textsuperscript{+} ions, incident at an angle of 5$^{\circ}$ to the samples surface; with sacttered ions detected at an angle of 175$^{\circ}$ to the direction of the primary beam. 20 random adjustments in the tilt of the sample, between +2$^{\circ}$ and -2$^{\circ}$, were made over the course of each measurement to average out the potential effect of ion channeling in the substrate. Fitting of the resulting RBS spectra was performed using the SIMNRA software package \cite{mayer2019ion}. ToF-ERDA measurements were performed using a primary beam of 36\,MeV \textsuperscript{127}I$^{8+}$ primary ions, incident at 67.5$^{\circ}$ with respect to the sample surface-normal and recoils were detected at an angle of 45$^{\circ}$ relative to the path of the primary beam. The ToF-ERDA detector system consisted of a time-of-flight telescope followed by a gas ionisation chamber, so that both the energy and time-of-flight of the recoiled particles were recorded in coincidence \cite{strom2016combined}. Composition depth profiles from ToF-ERDA measurements were calculated using the Potku software package \cite{arstila2014potku}. The recorded RBS and ToF-ERDA spectra were analyzed self-consistently following an iterative approach, in which the obtained information from ToF-ERDA analysis was used as boundary condition to fit the RBS spectra. Photoelectron Spectroscopy was utilized to identify the elemental composition and to investigate the electronic structure of the BaZrS$_3$ surfaces. The as-prepared BaZrS$_3$ sample had a composition of Ba: 19.1 \%, Zr: 21.5 \%, and S: 59.4 \%, as measured by RBS. Additionally, XPS was employed to determine the VBM positions with respect to the Fermi level for all the samples, allowing us to study the doping type of the samples before and after annealing. Since XPS measurements require good conductivity of the samples, we separately prepared BaZrS$_{3}$ films on a Si substrate and annealed them along with quartz samples. XPS measurements were acquired at the Kai-Siegbahn laboratory at Uppsala University, provided by two monochromatic sources and the hemispherical electron analyzer EW4000 (Scienta Omicron). The spectra were measured by Al K$\alpha$ (1.487 keV) and Ga K$\alpha$ (9.25 keV) X-rays, which allowed to probe different depths into the surface (Al K$\alpha$: $<$ 1 nm; Ga K$\alpha$ $\sim$ 10 nm). \textcolor{black}{Before the measurements, the samples were mounted on a carbon tape on the sample holder, together with a gold foil for the energy calibration of the spectra, and the conductivity of the samples was ensured by a multimeter.} During acquisition, the pressure of the main chamber was $<$ 1.0 $\times$ 10$^{-9}$ mbar. The core level and valence band spectra measured at Ga K$\alpha$ photon energy were collected with a pass energy of 300 eV and with an energy step size of 0.1 eV. \textcolor{black}{All the XPS spectra were energy calibrated to the Fermi level by measuring a Au foil and by} setting the binding energy position of Au 4f$_{7/2}$ at 84.0 eV. The peak position of the S 1s peak for all the samples was achieved by fitting the core level spectra following the least square error method, while the position of the valence band maximum for the as-prepared sample was determined by a linear intersection of the valence band edge.

\subsection{Results and Discussion}

\subsubsection{Optical Properties}

The as-prepared BaZrS$_3$ thin films appear orange in color on a transparent substrate. Interestingly, they changed to dark brown after the annealing (as shown in the inset of Fig.~\ref{optical}). An optical image for all four samples is provided in the supplementary information (Fig. S1), clearly demonstrating that the color of BaZrS$_3$ thin films darkens with increasing annealing duration or temperature. 
\begin{figure}[htbp]
\includegraphics[width=0.49\textwidth]{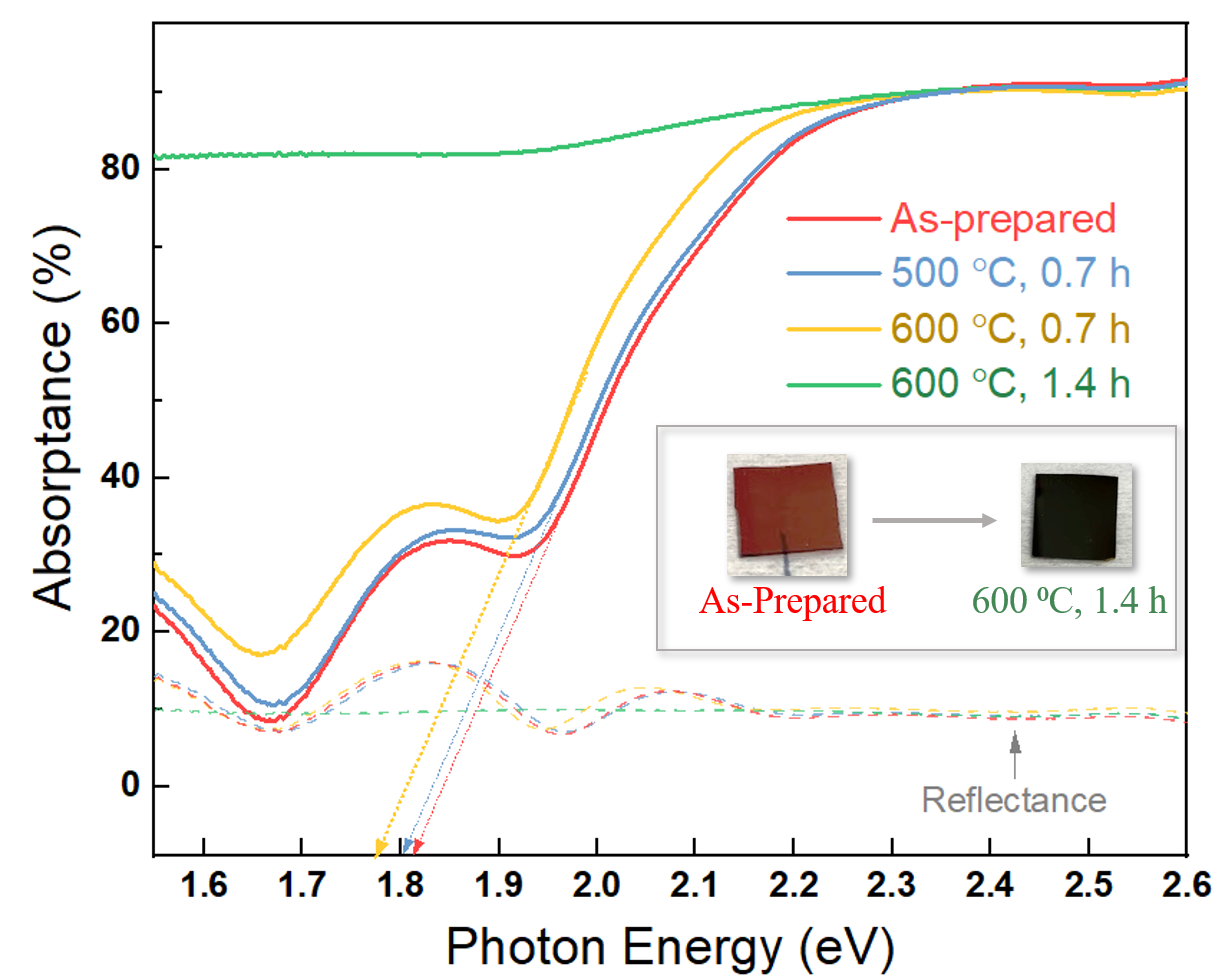}
\caption{Absorptance data (solid lines) of as-prepared and annealed BaZrS$_3$ thin films measured $\emph{via}$ UV-Vis spectroscopy in the range of 475 to 800 nm. Inset shows the optical image of the as-prepared and annealed sample.}
\label{optical}
\end{figure}
To understand this phenomenon, we measured the reflectance and transmittance of these films $\emph{via}$ using UV-vis spectroscopy. Fig.~\ref{optical} shows the absorptance (solid lines) of the annealed and as-prepared films. The absorptance edge of the as prepared film, 500-0.7 h, and 600-0.7 h films fall in the range of 1.82 to 1.77 eV. Importantly, the absorption edge shifts to the red region with increasing annealing temperature and duration. The sample annealed at 600-1.4 h exhibits very high absorptance for all wavelengths. These results are consistent with the optical images of these samples, where the film becomes darker after annealing and the sample 600-1.4 h appears dark brown. 

Bandgap shifts and associated color darkening with vacuum annealing are reportedly observed in metal oxides when annealed in a vacuum or oxygen-deficient environment \cite{dangi2023effect, wang2012oxygen, yang2017oxygen}. Generation of oxygen vacancies is the primary reason for such behavior, also altering electrical properties, as films become more conductive with annealing due to an increase in electron concentration with oxygen vacancy defects \cite{murauskas2022structure, kumar2021opto}. Given that all of the above observations are consistent with our findings on the BaZrS$_3$ thin films, we hypothesize that a similar phenomenon is occurring in these films. Annealing in vacuum causes sulfur atoms to escape from the sample, creating S-vacancy defects (V$_{\mathrm{S}}$) at increasing concentration as annealing temperature and duration are increased. The increased number density of sulfur vacancies leads to the delocalization of impurity states, and due to the shallow nature of defects it starts overlapping with the valence band edge (further discussed in next section). Consequently, a redshift of the absorption edge is observed in Fig.~\ref{optical}. Moreover, creation of V$_{\mathrm{S}}$ in the system should also manifest in the electrical properties as a S-vacancy defect is a positively ionized charged defect, inducing n-type of conductivity in the film. 

\subsubsection{Electrical properties}

\begin{figure*}[t!]
\centering
\includegraphics[width=0.98\textwidth]{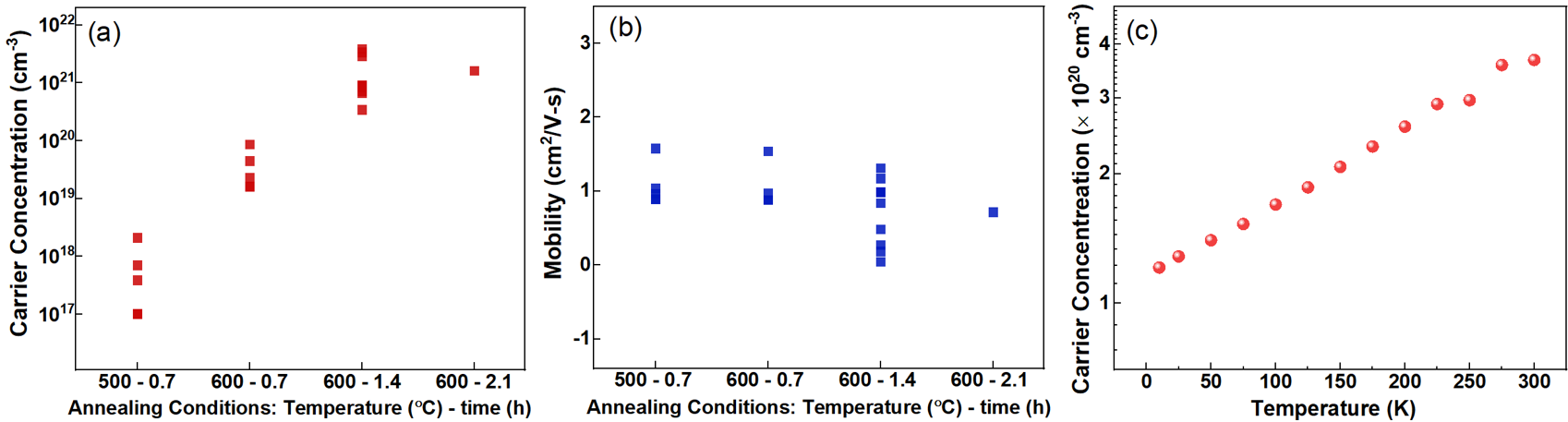}
\caption{Electrical properties of annealed BaZrS$_3$ thin films; (a) Carrier concentration and (b) Carrier mobility. (c) Semilogarithmic plot of temperature-dependent carrier concentration of the sample 600 $^{\circ}$C - 0.7 h in the range of 10 to 300 K.}
\label{elec}
\end{figure*}

Electrical properties of samples were \textcolor{black}{characterized} by Hall measurement, and contacts were made in Van der Pauw configuration with Ti/Au $\emph{via}$ sputtering. The ohmic nature of contacts was validated by a current sweep. All samples showed negative Hall co-efficient indicating n-type of conductivity in the film as hypothesized in the previous section. To verify this further, we also measured the Seebeck co-efficient of each sample (shown in Fig. S2). The results from both Seebeck and Hall measurements confirm n-type of conductivity in the films. Beyond determining the conductivity type, we also investigated how the conductivity changes with annealing parameters. Initially, our as-prepared BaZrS$_3$ film were highly resistive in nature. However, after annealing, conductivity (around 2 S/cm) was observed in the film. Fig.~\ref{elec} (a) shows the carrier concentration of the films, annealed under different conditions. To improve data reliability, we annealed multiple samples simultaneously in a single run, all diced from a single BaZrS$_3$ sample, and measured their electrical properties. The charge carrier concentration at different annealing conditions show a clear increase with increasing annealing temperature-duration. The carrier concentration directly correlates to the number density of defects that are ionized at room temperature. This implies that annealing creates positively charged defects (donor) in the samples, inducing n-type of conductivity in the films, and their number density increases with the annealing temperature-duration. It is noteworthy to mention that beyond a certain point, the carrier concentration saturates, approaching to metallic regime (10$^{22}$ cm$^{-3}$) (Fig S3). At the same time, the samples start looking dark brown, as observed in the optical image (Fig. S1).

The corresponding mobility of these films, however, did not show any dependence on the annealing temperature. Fig.~\ref{elec} (b) shows that the mobility values for all annealed samples appear around 1 cm$^2$V$^{-1}$s$^{-1}$ for all samples. While there might be some dependency of mobility on annealing conditions, such small values of mobility and the range of deviation, restrains further analysis. Considering that mobility remains unchanged with increasing carrier concentration, we propose that ionized impurity scattering does not dominate the charge transport at room temperature in our films. Rather, either grain boundary scattering or phonon scattering must be dominant at room temperature. Identifying the dominant scattering mechanism and improving the mobility are part of our ongoing work. Given the nm-sized grains in our film, we believe that improving the grain size is one direction for enhancing the carrier mobility.

We further analyzed the sample using temperature dependent Hall measurement to obtain characteristics of the charged defect and dominant scattering mechanism in the BaZrS$_3$ thin films. The semi-logarithmic graph in Fig.~\ref{elec} (c) illustrates the temperature-dependent carrier concentration of the sample 600-0.7, measured from 10 K to 300 K. The conductivity also increases monotonically with temperature (Fig. S4). As seen in Fig.~\ref{elec} (c), the concentration increases with temperature, a typical behavior observed in semiconductors \cite{blakemore2002semiconductor, aggarwal2019intrinsic, murphy2019minority}. However, the range of carrier concentration values on the Y-axis is rather narrow (1 $\times$ 10$^{20}$ to 4 $\times$ 10$^{20}$ cm$^{-3}$) even at 10 K, the concentration being still as high as 1 $\times$ 10$^{20}$ cm$^{-3}$. This indicates that even at 10 K, the defects are ionized to the order of 10$^{20}$ cm$^{-3}$. With such a high defect concentration, their wave functions overlap, and the defects can no longer be described as discrete localized levels within the bandgap \cite{blakemore2002semiconductor, pearson1949electrical, conwell1956impurity}. Instead they form an impurity band of non-localized states. Impurity metallic behavior is an extreme example of this situation, where the impurity band become contiguous to one of the principal bands. The present case is not so extreme, as evidenced by the fact that  the number of carriers is not constant throughout the measurement ranging from 10 to 300 K (Fig.~\ref{elec} (c)) \cite{blakemore2002semiconductor}. Using the Arrhenius equation at high temperatures, the thermal activation energy was determined, which yielded a value of $\sim$ 8 meV. Such small activation energy in the present case shows that the donor defect is a ''weak'' impurity metal state, i.e. the impurity band will consist of non-localized states all in the immediate neighborhood of the energy E$_C$ (CBM) and only marginally affects the conduction band \cite{blakemore2002semiconductor}. Similar defect studies of silicon and germanium show that the activation energy decreases with increasing defect number density and eventually goes down to zero, when it is contiguous to CB/VB \cite{pearson1949electrical, conwell1956impurity}. The existence of this impurity band near the CBM most likely contributes to the red shift in the absorption edge (Fig.~\ref{optical}).

The combined results from Fig.~\ref{optical}, Fig.~\ref{elec} (a), and Fig.~\ref{elec} (c) indicate that annealing BaZrS$_3$ creates defects and their number density increases with increasing annealing temperature/duration. These defects most likely are S-vacancies, given that the samples are annealed in a S-devoid atmosphere and sulfur's high vapor pressure favoring its transition to the vapor phase, unlike other elemental constituents of the material. Also, the temperature-dependent carrier concentration indicates a low activation energy value. This is consistent with existing theoretical research on BaZrS$_3$, which indicate that the positive ionization energy level associated with S-vacancy defects lies close to CBM in Fig.~\ref{defects} \cite{wu2021defect, thakur2023recent, wu2021defect, osei2021examining}. This behaviour is closely analogous to the behaviour of oxide perovskites treated under O-deficient conditions \cite{murauskas2022structure, kumar2021opto}. To further support our hypothesis, we estimated the chemical composition of the sample before and after annealing, which is discussed in the following section.

\subsubsection{Chemical Composition}

\begin{figure}
\centering
\includegraphics[width=0.4\textwidth]{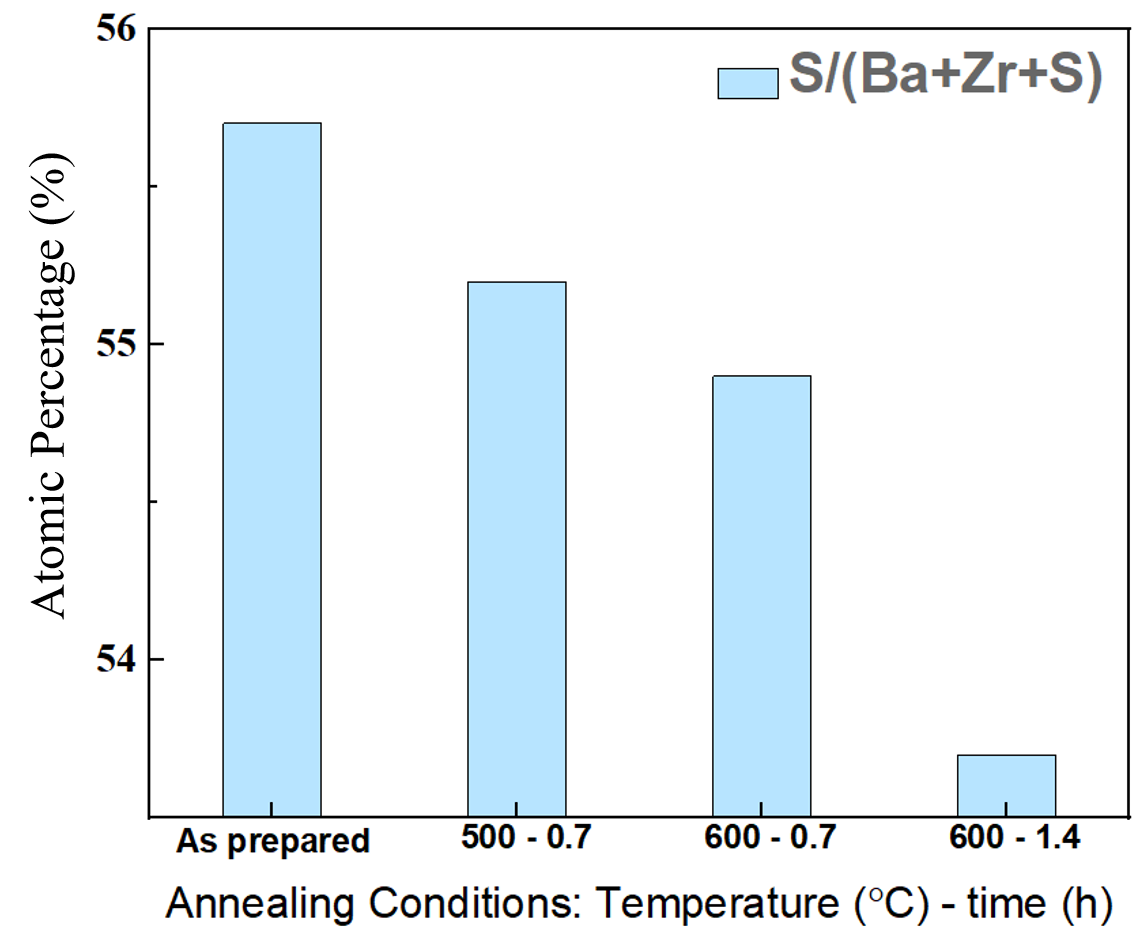}
\caption{Chemical composition of both as-prepared and annealed BaZrS$_3$ films, measured $\emph{via}$ surface energy dispersive spectroscopy (EDS).}
\label{erda}
\end{figure}

Fig.~\ref{erda} compares the atomic percentage of sulfur in both as-prepared and annealed BaZrS$_3$ films using surface energy dispersive spectroscopy (EDS). In EDS, the measurement is conducted from the top of the sample, and the information is more sensitive to the surface although considers the entire film thickness and partial depth of the substrate. Importantly, we observe a decrease of around 2 \% in the atomic percentage of sulfur after annealing. This indicates that the annealed BaZrS$_3$ samples exhibit higher S-deficiency than the as-prepared BaZrS$_3$. 

Considering the density of BaZrS$_3$ to be 4.22 gm/cm$^3$, a volume of 1 cm$^3$ corresponds to 0.078 $\times$ 10$^{23}$ molecules \cite{comparotto2022synthesis}, which translates to 2.35 $\times$ 10$^{22}$ S-atoms/cm$^3$. Based on this, a reduction of S by 1 \% leads to an electron concentration of around 4.6 $\times$ 10$^{20}$ cm$^{-3}$ and 2 \% to 1.17 $\times$ 10$^{21}$ cm$^{-3}$ (SI). Here, the assumption is that all S-vacancy defects are completely ionized (V$_{\mathrm{S}}^{\bullet  \bullet}$) with each defect donating 2 electrons to the conduction band. These calculated values closely match with the observed values in Hall measurement, as shown in Fig.~\ref{elec} (a). Although we also note that an overall  change of 1-2 \% in S content is difficult to measure accurately, especially due to the possibility of a concentration gradient with depth (we expect most S loss from the surface), and the fact that some S-loss could be compensated by inclusion of oxygen atoms. The composition measurements alone thus cannot prove the existence of V$_{\mathrm{S}}$. Nevertheless in the overall picture, we consider creation of V$_{\mathrm{S}}$ to be the most consistent explanation for the observations, in particular the fact that n-type defects can be induced by vacuum annealing.

\subsubsection{XPS Analysis}
The chemical composition of the BaZrS$_{3}$ surfaces were further investigated using XPS at two different energies: a more surface-sensitive Al K$\alpha$ (1.487 keV) and Ga K$\alpha$ (9.25 keV), where the information depth reaches tens of nanometers. While the survey scans acquired at both energies (Al K$\alpha$ and Ga K$\alpha$) reveal all characteristic core levels of constituent atoms Ba, Zr, and S of the BaZrS$_{3}$ compound (See Fig. S5 and Fig. S6), the Al K$\alpha$ measurements also show additional peaks from Sn, O, and C (Fig. 5). These signatures are related to surface impurities formed during material synthesis and surface oxidation upon subsequent exposure to air. In particular, the peaks related to Sn for the as-prepared and 500-0.7h samples start disappearing when the annealing temperature was raised to 600 $^{\circ}$C, indicating that we reached \textcolor{black}{the evaporation} temperature of Sn-related compounds in vacuum (10$^{-9}$ mbar). On the other hand, the surveys for the samples 600-0.7h and 600-1.4h detect an unknown impurity species at approx 305 eV binding energy (Fig. S6), probably coming from the sulfurization furnace. The absence of these impurity peaks in the Ga K$\alpha$ survey for all the samples establishes that these impurities must confined within the first few nanometers of the surface, and are unlikely to affect the bulk properties of the films. The annealing process also visibly affects chemical bonding of S atoms in the films. The Al K$\alpha$ XPS measurements reveal peaks corresponding to oxidized S-species, except for the sample 600-1.4h where these contributions disappear in favour of the lowest binding energy S peak (Fig. S7), identified previously as the S peak of BaZrS$_{3}$ \cite{riva2024electronic, mukherjee2023interplay}. The measurements using Ga K$\alpha$, on the other hand, detect \textcolor{black}{mainly} the perovskite S peak, significantly lower O content, and no adventitious C 1s peak. Moreover, for Ga K$\alpha$, the peaks related to Sn visible for the as-prepared and 500-0.7h samples, disappeared as annealing temperature increases to 600 $^{\circ}$C, similar to our observation from Al K$\alpha$. All these observations indicate that the annealing process affects surface chemistry at up to least 10 nm from the surface, while the bulk of the films is primarily comprised of the BaZrS$_{3}$ perovskite. \textcolor{black}{Nonetheless, from our XPS analysis, we can conclude that the annealing process in vacuum has the additional advantage of improving the surface quality, together with improving electrical properties (see above). }

\subsubsection{Band diagram}

\begin{figure}[htbp]
\centering
\includegraphics[width=0.49\textwidth]{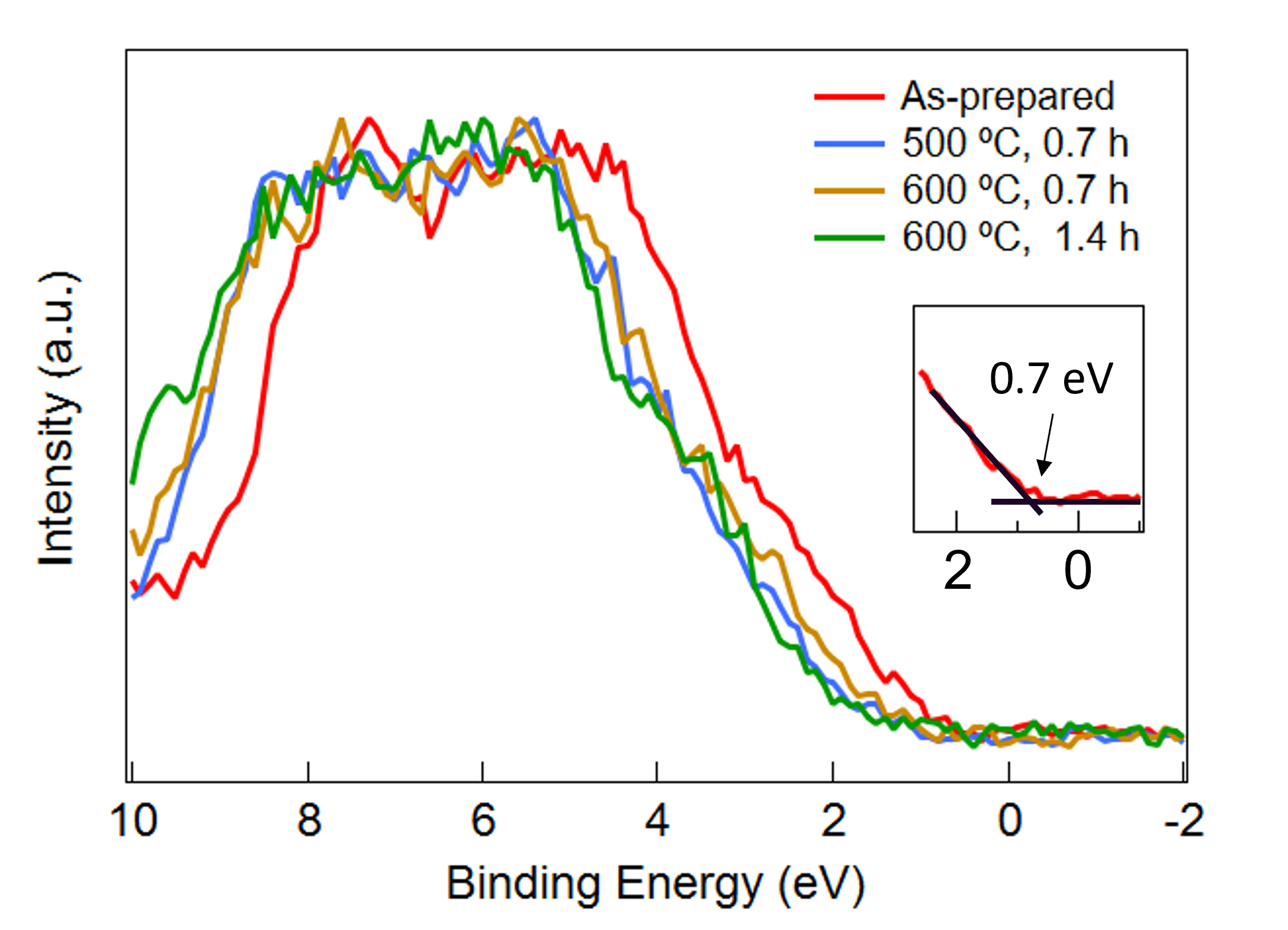}
\caption{Valence bands of both as-prepared (green) and annealed BaZrS$_3$ films, measured $\emph{via}$ X-ray photoelectron spectroscopy with photon energy 9.25 keV. The spectra are energy calibrated by setting Au 4f$_{7/2}$ at 84.0 eV. The intensity is normalized for the ease of comparison. Inset: zoom on the valence band edge of the as prepared sample for the determination of the valence band maximum.}
\label{val}
\end{figure}

\begin{figure}[b]
\centering
\includegraphics[width=0.49\textwidth]{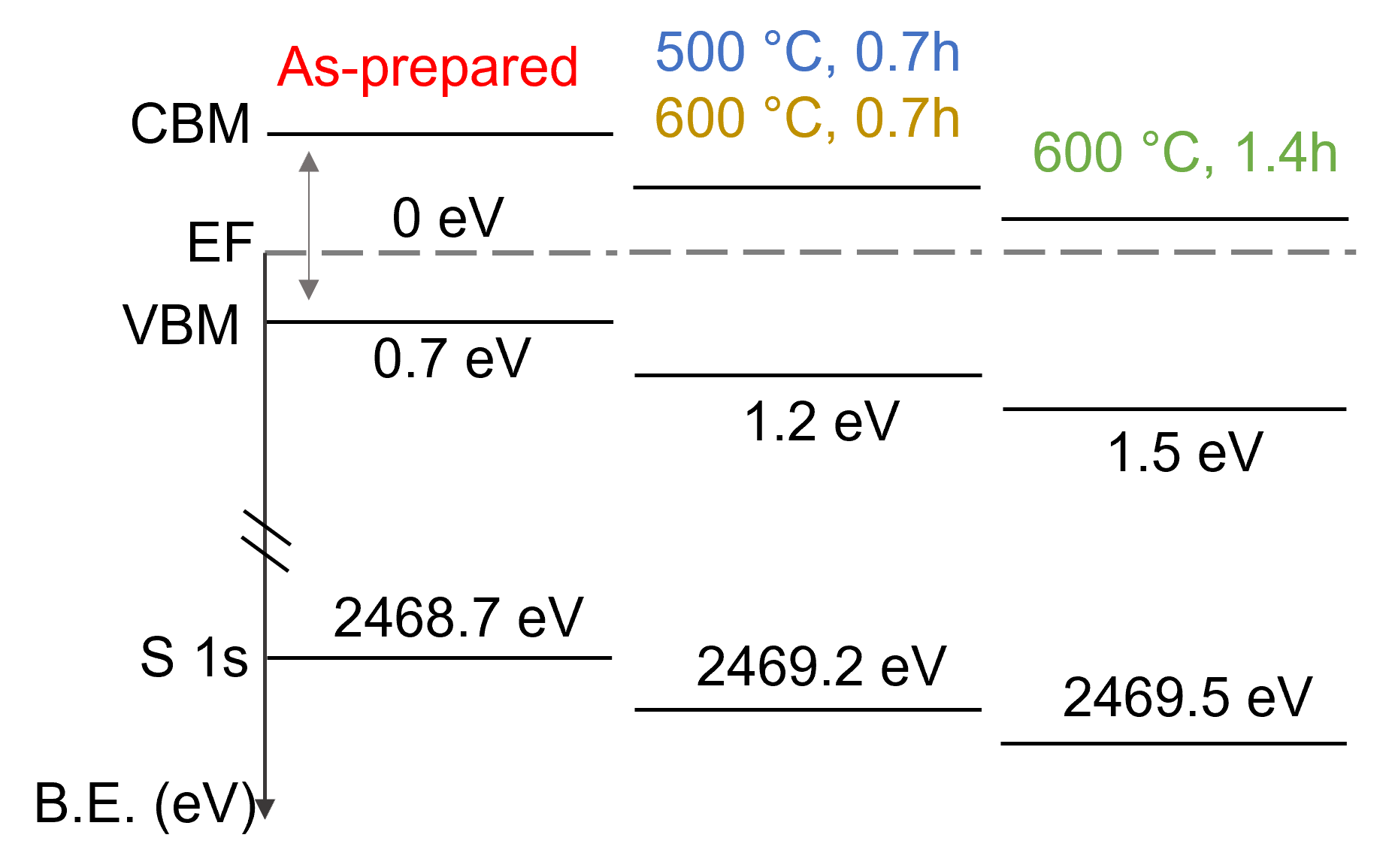}
\caption{Energy diagram of the BaZrS$_3$ samples: as-prepared, 500-0.7 h, 600-0.7h and 600-1.4h, as derived from the XPS analysis. The binding energy scale starts from the Fermi level at 0 eV, as from the spectroscopy measurements calibrated vs Au 4f$_{7/2}$. The utilized band gap value is 1.8 eV for all samples.}
\label{diagram}
\end{figure}

The  valence band spectra acquired with Ga K$\alpha$ was utilized to determine the energy shift of the valence band maxima (VBM) with respect to the Fermi level \textcolor{black}{(see Experimental Section)}. Fig.~\ref{val} compares the valence band spectra of the as-prepared and annealed BaZrS$_{3}$. We find that upon annealing, valence band edge positions shift towards higher binding energy values with respect to the Fermi level. These energy offsets characteristic to the annealing conditions are related to the nature of doping \textcolor{black}{and} modify the energy diagram of the BaZrS$_{3}$ compounds (Fig.~\ref{diagram}), and consequently changing in conductivity behavior in these samples. \textcolor{black}{For the comparison of how the VBM shift upon annealing for the sample series, we used the intersection point of a linear fit to the valence band edge. Following this method we found the VBM appearing at 0.7 eV vs Fermi level for the as-prepared BaZrS$_{3}$ sample, as already found in our previous work \cite{riva2024electronic}}. An identical offset as for the valence band is observable in the position of the S 1s peak (see Fig. S8). From the fit of the S 1s peaks for the samples, we could \textcolor{black}{more} precisely determine the amount of shift of the valence band maxima between the samples. For the as-prepared sample, the main S 1s peak position is at 2468.7eV. When the samples are annealed at 500-0.7 h and 600-0.7 h, the main S 1s peak position appears at 2469.2 eV. Finally, the sample annealed at 600-1.4 h has a S 1s main peak at 2469.5 eV, with an increase of binding energy of 0.7 eV compared to the as-prepared sample. Considering the band gap of 1.8 eV, we can identify the conduction band minimum position, which allows us to compile a relative band diagram for all four samples in Fig.~\ref{diagram} \cite{han2023p, filippone2020discovery, sun2015chalcogenide}.  
From the figure, one can conclude that annealing causes the Fermi level to shift closer to the conduction band, thereby providing the n-type of conductivity in  BaZrS$_{3}$ thin films. Our analysis of the VB reveals how annealing temperature or annealing time can be utilized as critical design parameters to control transport properties of BaZrS$_{3}$ thin films.

The position of the Fermi level above the midgap region (E$_{\mathrm{g}}$/2) in annealed samples corroborates the observed n-type of conductivity in both Hall and Seebeck coefficient results. The shift of the Fermi level towards CBM in Fig.~\ref{diagram} is consistent with Hall measurement findings. Increasing annealing temperature-duration, as shown in Fig.~\ref{elec} (a), leads to higher carrier concentration and a corresponding rise in the Fermi level towards the CBM. It is noteworthy that in the as-prepared sample (Fig.~\ref{diagram}), the Fermi level lies below the midgap region (E$_{\mathrm{g}}$/2), suggesting p-type conduction. The extremely high resistivity of the as-prepared films hindered accurate electrical measurements to make definitive claims about defects. However, considering our synthesis conditions, subsequent composition analysis, and theoretical studies (Fig.~\ref{defects}), we believe that the dominant acceptor type of defects mostly include V$_{\mathrm{Ba}}$ and/or S$_{\mathrm{i}}$ defects. Nevertheless, a systematic study is required to further validate these assertions. Given our hypothesis regarding dopant-induced impurity band, a precise determination of the Fermi level position through electrical data and in-direct comparison with the XPS values, becomes challenging. However, the qualitative agreement between the XPS findings and the electrical data provides a novel approach, correlating structural analysis with functional properties.

\subsection{Summary}

To summarize, we investigated the effect of annealing BaZrS$_3$ thin films in a S-devoid (high vacuum) environment and how it impacts transport properties. We find that vacuum annealing induces S-vacancy (V$_{\mathrm{S}}$) defects, with their ionization level lying close to the conduction band minimum. Consequently, the initially insulating BaZrS$_3$ films exhibit n-type conductivity, with a significant increase of up to 2 S/cm. Notably, the number density of V$_{\mathrm{S}}$ defects demonstrates a direct correlation with the annealing temperature and duration. Compositional analysis $\emph{via}$ EDS shows that the S atomic percentage reduces by more than 2 \% upon annealing, resulting in a charge carrier concentration of $\approx$ 1 $\times$ 10$^{21}$ cm$^{-3}$ at room temperature, as established from Hall measurement. The emergence of n-type of conductivity post-annealing is also evident from the XPS results, where the Fermi level, initially positioned \textcolor{black}{around} the midgap (E$_{\mathrm{g}}$/2) region in as-prepared film, starts shifting towards the conduction band minimum with increasing annealing temperature and duration. The findings of this study underscore the ability of subjecting BaZrS$_3$ to a S-poor atmosphere during annealing to create S-vacancy defects, transforming it into an n-type semiconductor. The mobility in our films remains modest ($\sim$ 1 cm$^2$V$^{-1}$s$^{-1}$), and further optimization of the synthesis process is necessary to enhance its performance, particularly for applications in solar cells.

\vspace{-5pt}

\begin{acknowledgments}

We thank Pedro Berastegui for his help with the vacuum annealing furnace and Jan Keller for his help with UV-Vis measurements. The authors acknowledge CA21148 STSM Exchange Visit Grant by COST European Cooperation In Science \& Technology that enabled collaborative work with Prof. Monica's group at University of Twente, NL. The authors gratefully acknowledge the Göran Gustafsson Foundation (grant no. 1927), the Swedish Research Council (2017-04336, 2023-05072, 2018-04330, and 2018-06465), the Swedish Energy Agency (P50626-1), and the STANDUP Strategic Research Area for Energy to finance this research. Infrastructural grants by VR-RFI (grant numbers 2017-00646--9, 2019--00191) and SSF (contract RIF14-0053) supporting operation of the accelerator employed for ion beam analysis are gratefully acknowledged. 

\end{acknowledgments}.
%
\end{document}


\begin{center}
     \underline{\Large{Supplementary Information}}
\end{center}
\title{
Charge Transport and Defects in Sulfur-Deficient Chalcogenide Perovskite BaZrS$_3$
}
\author{Garima Aggarwal$^{a\dag}$}
\email{garima.agrawalsk@gmail.com\\ $\dag$Now at ReNew Private Limited, Gurgaon - India\\$^{\ddag}$These authors contributed equally}

\author{Adeem Saeed Mirza$^{b\ddag}$, Stefania Riva$^{c\ddag}$}
\author{Corrado Comparotto$^a$, Robert J. W. Frost$^d$, Soham Mukherjee$^c$, Monica Morales-Masis$^b$, H$\mathring{a}$kan Rensmo$^c$, and Jonathan Staaf Scragg$^a$}

\affiliation{$^a$Division of Solar Cell Technology, Department of Materials Science and Engineering, Uppsala University, Uppsala 75237, Sweden\\
$^b$ MESA+ Institute for Nanotechnology, University of Twente, Enschede 7500 AE, The Netherlands\\
$^c$ Molecular and Condensed Matter Physics, Department of Physics and Astronomy, Uppsala University Box 516, SE-75120 Uppsala, Sweden\\
$^d$ Division of Applied Nuclear Physics, Uppsala University, Uppsala 75237, Sweden}

\maketitle


\newpage

\begin{figure}
\centering
\includegraphics[width=0.7\textwidth]{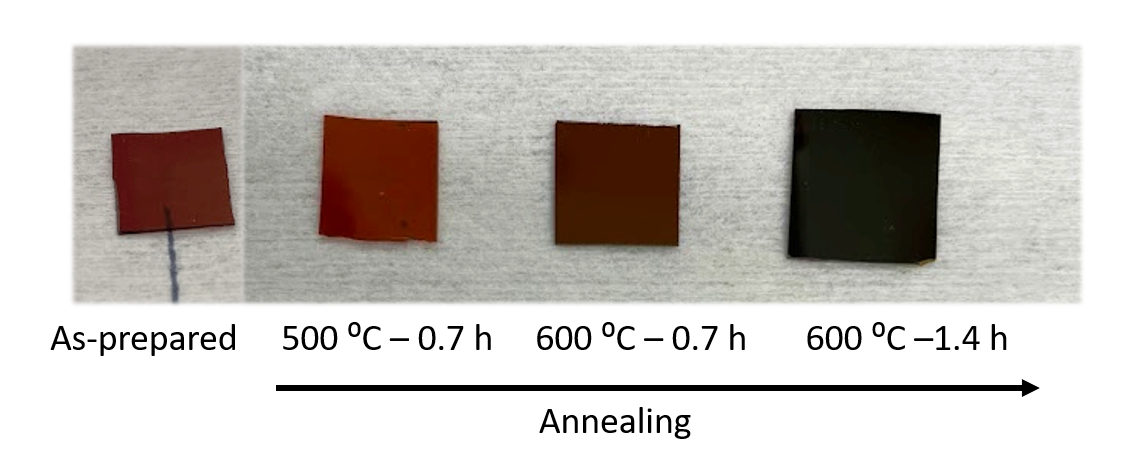}
\caption{Optical images of BaZrS3 films on quartz substrate; as prepared and annealed in high vacuum.}
\end{figure}
\begin{figure}
\centering
\includegraphics[width=0.98\textwidth]{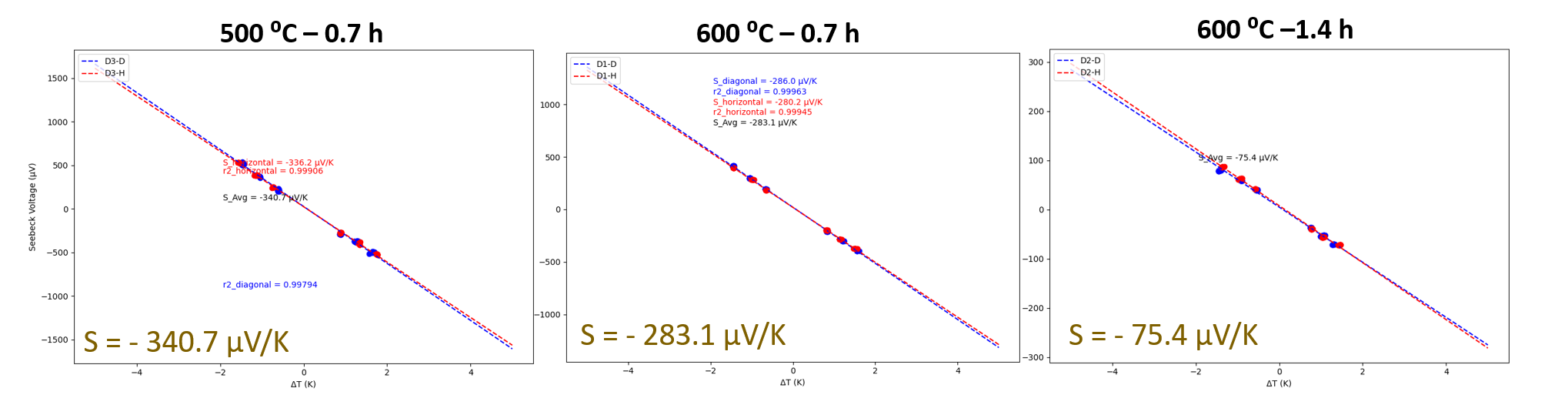}
\caption{Seebeck coefficient of annealed BaZrS$_3$ thin films. The negative coefficient indicates n-type of conductivity in the films, and the decreasing absolute value from left to right indicates increasing carrier concentration.}
\end{figure}
\begin{figure}
\centering
\includegraphics[width=0.6\textwidth]{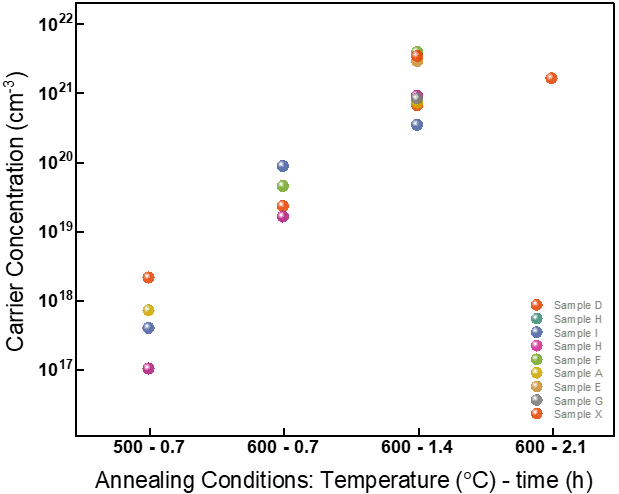}
\caption{Carrier concentration of both as-prepared and annealed BaZrS$_3$ thin films. There is a saturation observed after 600 $^{\circ}$C-1.4 h annealing, therefore, results are presented in the manuscript for first three annealing conditions; 500-0.7, 600-0.7, and 600-1.4.}
\end{figure}
\begin{figure}
\centering
\includegraphics[width=0.6\textwidth]{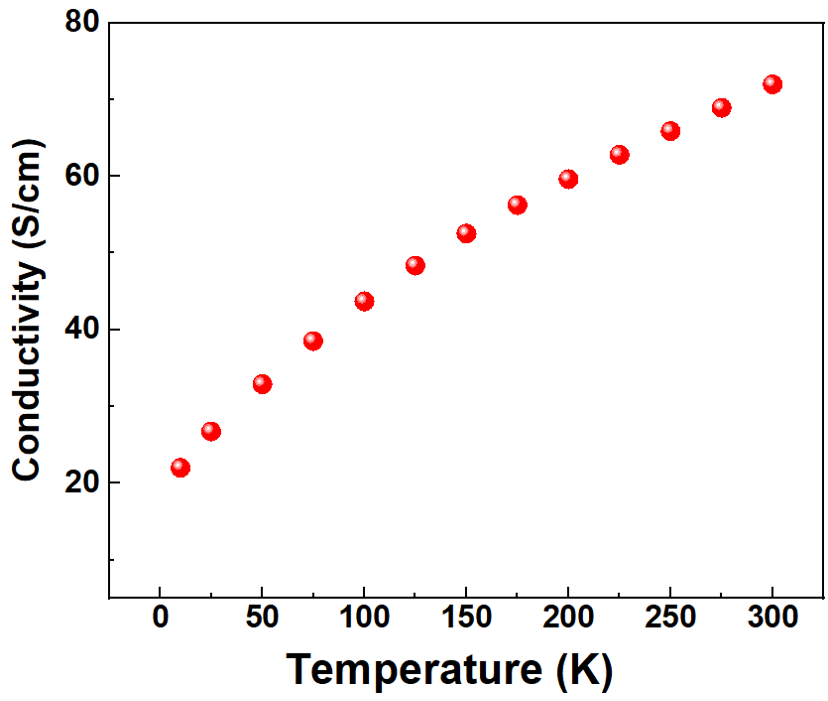}
\caption{Temperature dependent conductivity of BaZrS$_3$ thin film annealed at 600-0.7.}
\end{figure}

\begin{figure}
\centering
\includegraphics[width=0.65\textwidth]{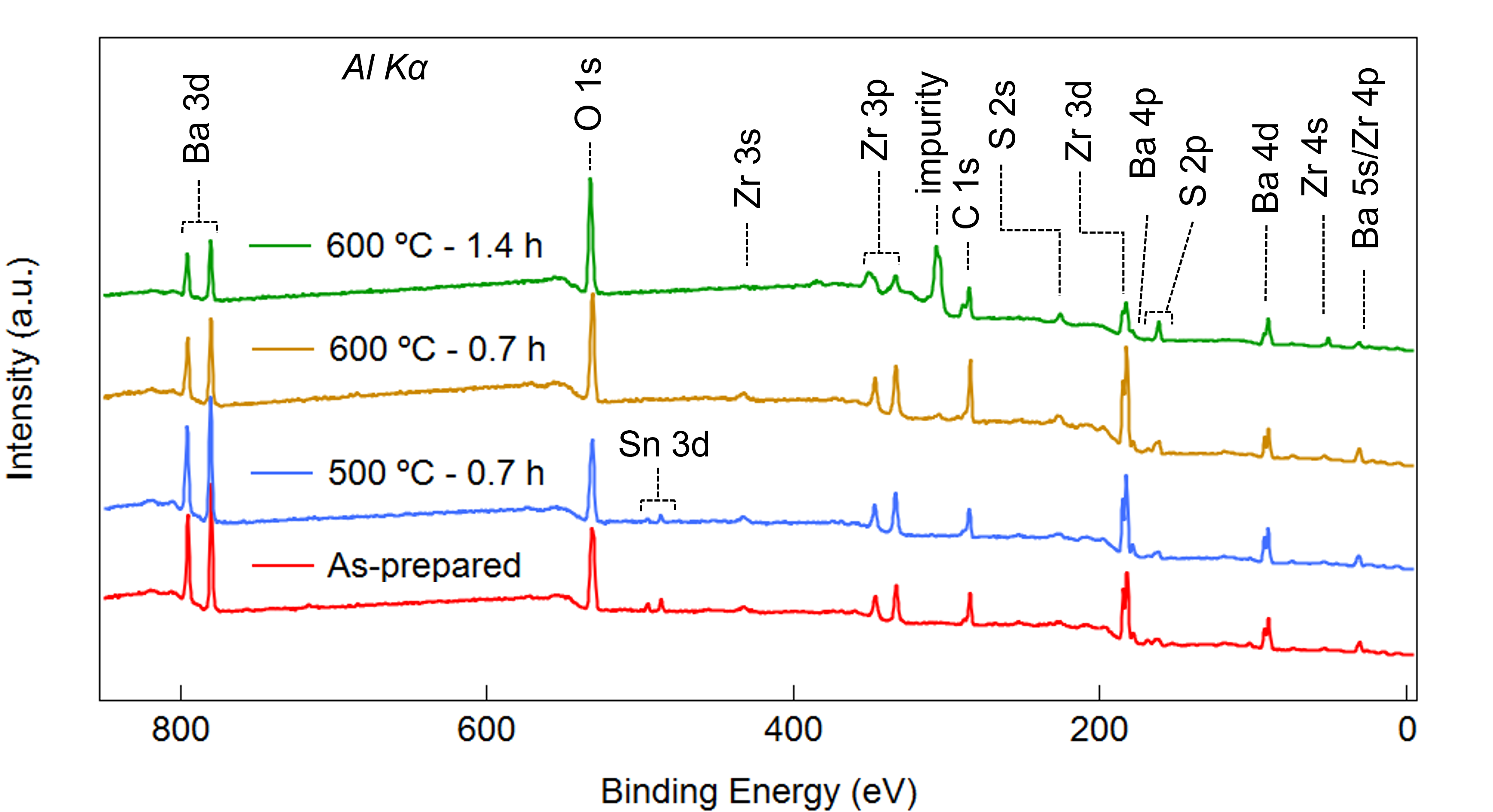}
\caption{Photoelectron spectroscopy survey scan of as-prepared and annealed BaZrS$_3$ thin films measured with Al K${\alpha}$ (1.487 keV), showing different core-level peaks indexed. Together with the BaZrS$_3$ characteristic peaks, the survey scans of 600-0.7 and 600-1.4 presents a peak around 307 eV, labelled as impurity. }
\end{figure}
\vspace{-5pt}
\begin{figure}
\centering
\includegraphics[width=0.4\textwidth]{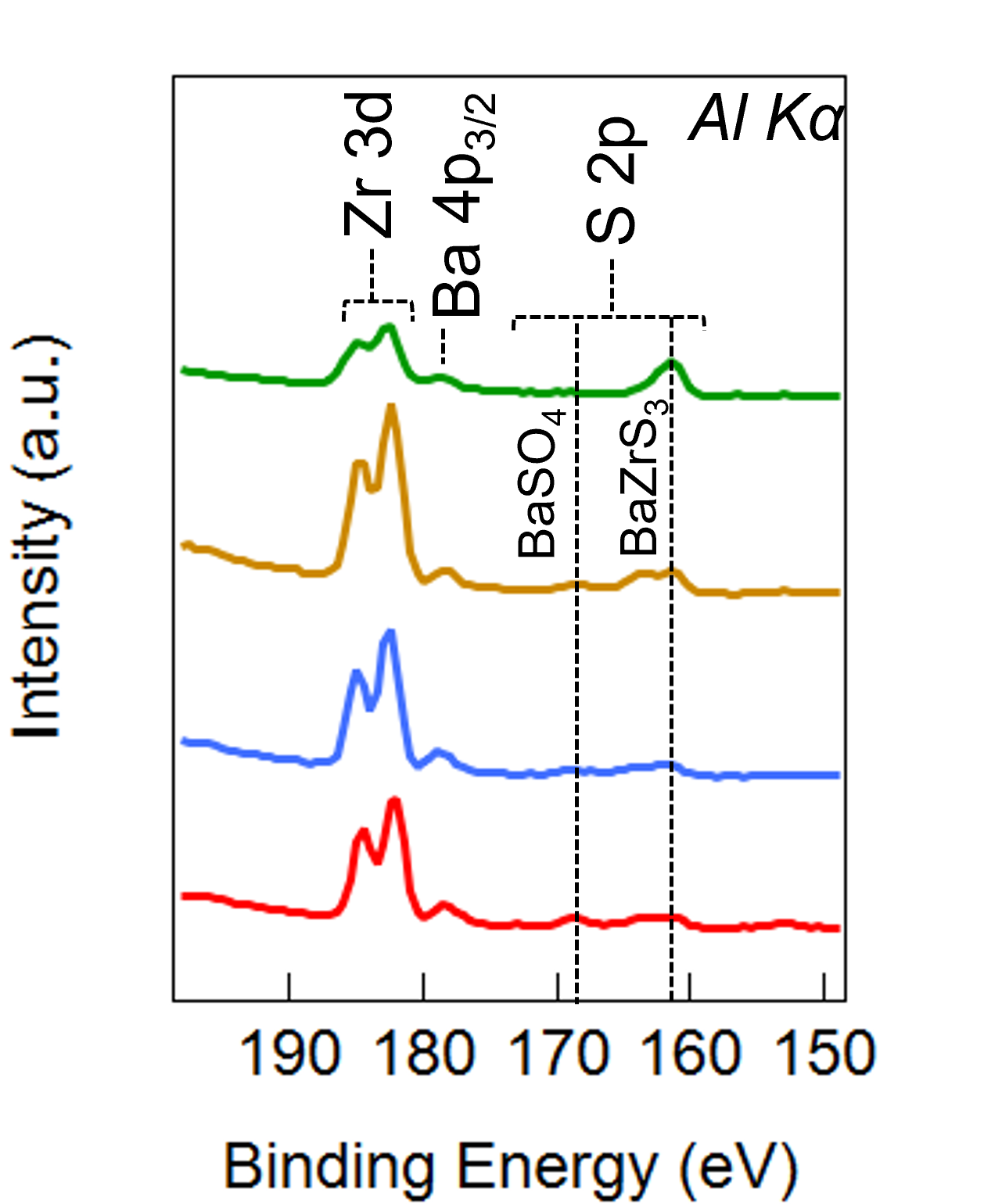}
\caption{Zoom into the region 150 - 200 eV of the photoelectron spectroscopy survey scan of as-prepared and annealed BaZrS$_3$ thin films measured with Al K${\alpha}$ (1.487 keV). The region contains the Zr 3d, Ba 4p$_{3/2}$ and S 2p peaks. The color code is the same as in Fig. S5 and Fig. S7. The as-prepared sample (red) and the thin films annealed at 500 $^{\circ}$C (blue) and at 600 $^{\circ}$C for 0.7h (yellow) show the S-O bond together with the S 2p from BaZrS$_3$, while the film annealed at 600 $^{\circ}$C for 1.4h presents only the BaZrS$_3$ S 2p peak. }
\end{figure}
\begin{figure}
\centering
\includegraphics[width=0.65\textwidth]{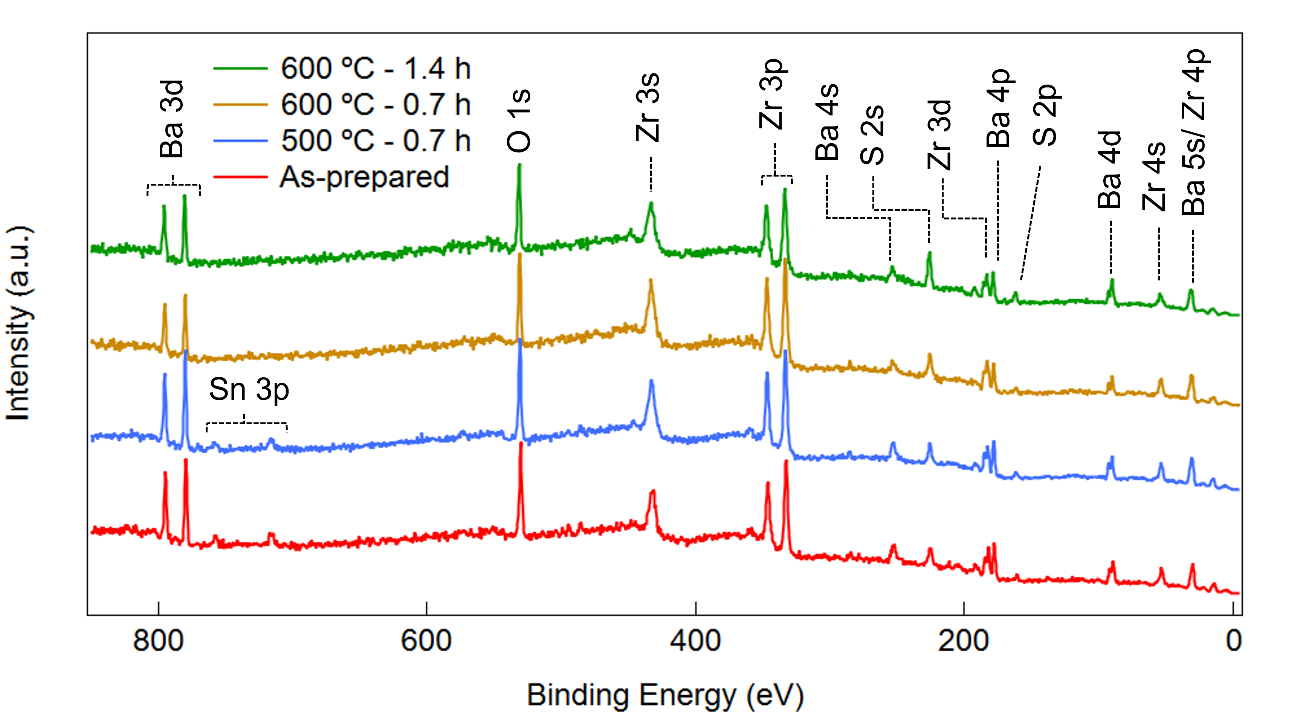}
\caption{Photoelectron spectroscopy survey scan of as-prepared and annealed BaZrS$_3$ thin films measured with Ga K${\alpha}$ (9.25 keV), showing different core-level peaks indexed.}
\end{figure}
\begin{figure}
\centering
\includegraphics[width=0.6\textwidth]{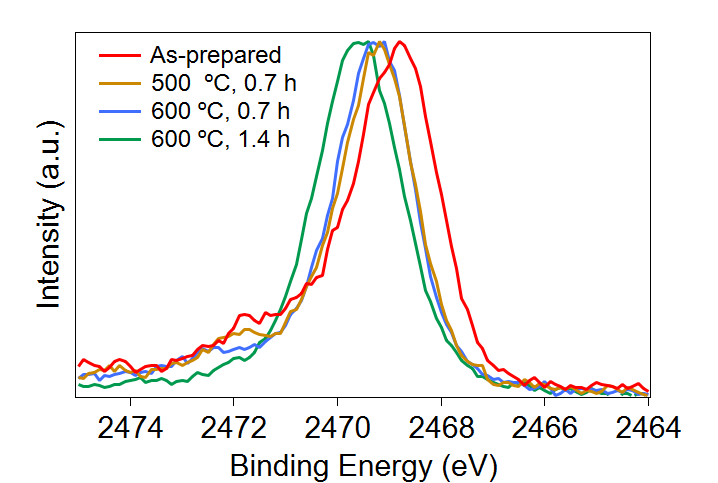}
\caption{S 1s peaks of as-prepared (green) and annealed BaZrS$_3$ thin films, measured via X-ray photoelectron spectroscopy with photon energy 9.25 keV. The spectra are energy calibrated by setting Au 4f$_{7/2}$ at 84.0 eV. The intensity is normalized for the ease of comparison. A clear shift to higher binding energy with annealing is observed.}
\end{figure}

\newpage

\begin{figure}
\underline{\large{Calculation of Electron Concentration in BaZrS$_3$ with S-vacancy defects}}

\vspace{8pt}

\centering
\includegraphics[width=0.97\textwidth]{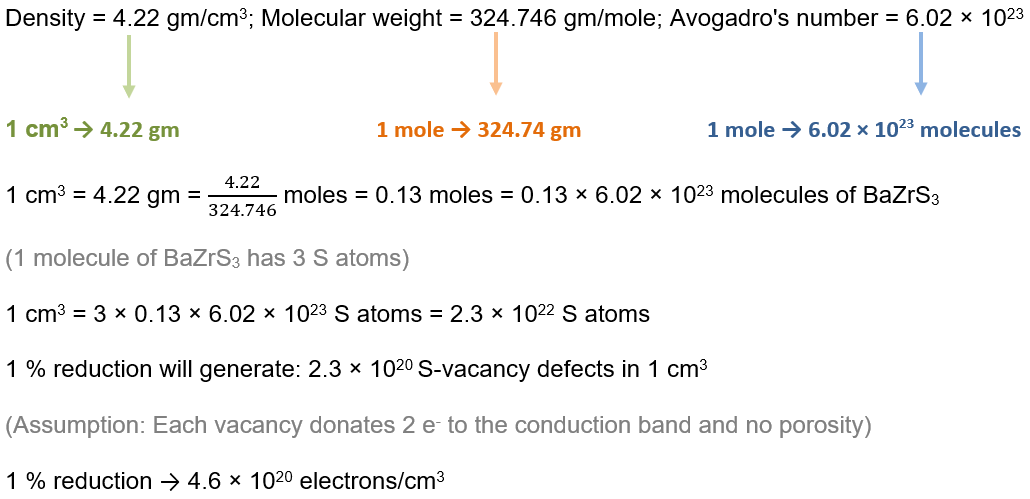}

\end{figure}